\documentstyle[]{article}
\font\cero=cmss10 scaled 1728 \font\uno=cmssbx10 scaled 1200
\begin{document}
\small{
\begin{flushleft}
{\cero  Simplectic geometry and the canonical variables for Dirac-Nambu-Goto and Gauss-Bonnet system in string theory} \\[3em]
\end{flushleft}
{\sf Alberto Escalante}\\
{\it  Departamento de F\'{\i}sica, Centro de Investigaci\'on y de Estudios Avanzados del I.P.N., \\
Apdo Postal 14-740, 07000 M\'exico, D. F., M\'exico. \\ 
Instituto de F\'{\i}sica, Universidad Aut\'onoma de Puebla,
Apartado postal J-48 72570, Puebla Pue., M\'exico,}
(aescalante@fis.cinvestav.mx) \\[4em]
\noindent{\uno Abstract} \vspace{.5cm}\\
Using a strongly covariant formalism given by Carter for the
deformations dynamics of p-branes in a curved background and a
covariant and gauge invariant geometric structure constructed on
the corresponding Witten's phase space, we identify the canonical
variables for Dirac-Nambu-Goto [DNG] and Gauss-Bonnet [GB] system
in string theory. Future extensions of the present results are
outlined.
\noindent \\

\begin{center}
{\uno I. INTRODUCTION}
\end{center}
\vspace{1em} \ The interest in physical systems characterized by
extended structures goes back to the XIX th Century and to Lord
Kelvin's "aether atoms'', for which a spatial extension was
postulated in order to accommodate a complex structure which would
be have both as
 an elastic solid (conveying the transverse wave motion of electromagnetism)
 and viscuss liquid (dragged by the earth in its orbital motion).\\
In the XX the Century, there have been three active motivations
leading to either classical or quantum extendons. On the other
hand, the physics of condensed matter  (including biological
systems) have revelated that membranes and two-dimensional layers
play an important role; in some case, there also appear
one-dimensional filaments (or strings). Similar structures appear
in astrophysics and cosmology, one example being the physics of
Black holes, in which the "membrane" is the  boundary layer
between the hole and the embedding spacetime, and another example
is
represented by the hypothetical cosmic strings.\\
In last years a considerable amount of effort has been devoted for
developing a quantum field theory of such extended objets (which
in fact, will constitute the ultimate framework) for a complete M
theory; however, it has not yet been fully developed. The problem
is that the dynamics of extended objets is highly nonlinear and
the standar methods are not directly applied. However, using a
covariant canonical formalism introduced by Witten \cite{1} in
recent letters \cite{2,3,4,5} the basic elements to quantize
extended objects (in particular bosonic p-branes) has been
explore, for example, in \cite{3} we established  the bases to
study the quantization aspects of p-branes with thickness,
because, when adding it to the [DNG] action has an important
effect in QCD \cite{6}, among other things. In \cite{4} has been
demonstrated that the presence of Gauss-Bonnet [GB] topological
term in the [DNG] action describing strings, has a dramatic effect
on the covariant phase space formulation of the theory, in this
manner, we shall obtained a completely different quantum field
theory. Recently, using the results given in \cite{4} we identify
the covariant canonical variables for [DNG] p-branes and [GB]
strings, among other things \cite{7}. However, we found a little
problem, because,  the canonical variables found for [DNG] are
identified with spacetime indices, whereas, the canonical
variables for [GB] strings with worldsheet indices, in this
manner, if we add the [GB] term to the [DNG] action describing
strings, we need identify the pullback on the canonical variables
for [GB] strings in order to obtain it in terms of spacetime
indices, and thus, to study in a covariant form the quantization
aspects for
[DNG-GB] system in string theory, but it was not clarified.\\
In this manner, the purpose of this article is to make, first,  a
generalization of the results presented in \cite{3,4} for a
general Lagrangian constructed locally from the geometry of the
worldvolume in a arbitrary background, after that, using a
strongly covariant formalism given by Carter \cite{8},  we
identify the canonical variables for [DNG-GB] system in string
theory, in this manner, we resolve
the problem found in \cite{7}.  \\
This paper is organized as follows. In Sect.II, we make a
generalization of the method utilized in \cite{3,4} for  a
Lagrangian constructed from the geometry of the worldvolume
embedding in a arbitrary background, confirming as special case
the results found in \cite{3}. In Sect.III, we make an outline of
the results found in \cite{9}, which, will be important the
developed of this paper. In Sect. IV  using a strongly covariant
scheme  of deformations introduced by Carter \cite{8}, we found
the canonical variables for [GB] system in string theory, that
unlike \cite{7}, the canonical variables of [GB] has spacetime
indices, which will be determinant for the treatment of [DNG-GB]
strings. In Sect. V with the results found in the previous
sections we identify the canonical variables for [DNG-GB] system
in string theory, and with this result,  we clarified the problem
that we found in \cite{7}. In Sect. VI we give the conclusions and prospects.\\

\setcounter{equation}{0} \label{c2}.

\noindent \textbf{II. Symplectic potentials for p-branes in a curved background}\\[1ex]
In recent letters, a covariant and gauge invariant symplectic
structure for [DNG] p-branes \cite{2}, for membranes with
quadratic terms in the extrinsic curvature \cite{3}, and for the
Gauss-Bonnet topological term propagating in a curved background
\cite{4} has been constructed. The form of constructing this
geometric structure is by means of identifying from the arguments
of the total divergences at the level of the Lagrangian a
symplectic potential that does not contribute locally to the
dynamics of the system, but its variation (the exterior derivative
on the phase space) generates a geometric structure. In this
manner, the purpose of this section is to generalize these results
for a Lagrangian constructed from the
geometry of the worldvolume embedding in a arbitrary spacetime.\\
For our aims, we will consider a local action depending on the
embedding functions $X^{\mu}$ which is both invariant under
worldvolume reparametrization  and under rotations of the normals
given by
\begin{equation}
S[X]= \int \sqrt{- \gamma} L d^{D}\xi,
\end{equation}
where the Lagrangian $L$ will be constructed locally from the
geometry of the worldvolume  as
\begin{equation}
L(\gamma^{ab}, K{_{ab}}^{i}, \widetilde \nabla_{a} K{_{bc}}^{i}),
\end{equation}
here, $\gamma^{ab}$, $ K{_{ab}}^{i}$ and $\widetilde \nabla_{a}$
is the metric induced, the extrinsic curvature and the covariant
derivative under rotation of the normal vector field  respectively \cite{10}.\\
Now, we need calculate the deformation of the Lagrangian given in
equation (2) to identify the equations of motion and the
symplectic potential for the theory described by the action (1).
For this, we descompose an arbitrary infinitesimal deformation of
the embedding $\delta X^{\mu}$ into its parts tangential and
normal to the worldvolume, this is
\begin{equation}
\delta X^{\mu}= e{^{\mu}}_{a} \phi^{a}+ n{^{\mu}}_{i} \phi^{i},
\end{equation}
where $n{^{\mu}}_{i}$ are the vector fields normal and
$e{^{\mu}}_{a} $ are the vector field tangent to worldvolume,
thus, the deformation operator is defined as
\begin{equation}
D=D_{\delta} + D_{\Delta},
\end{equation}
where
\begin{equation}
D_{\delta}= \delta^{\mu}D_{\mu}, \quad \quad \delta^{\mu}=
n{_{i}}^{\mu} \phi^{i},
\end{equation}
and
\begin{equation}
D_{\Delta}= \Delta^{\mu} D_{\mu}, \quad \quad \Delta^{\mu}=
e{_{a}}^{\mu} \phi^{a},
\end{equation}
in this manner, the variation of the equation (1) with the
Lagrangian (2) is given by
\begin{equation}
\delta S  =  \int \sqrt{-\gamma} \nabla_{a}(L \phi^{a})d^{D} \xi +
\int \sqrt{-\gamma} [ K^{i}\phi_{i}L + H{^{ab}}_{i} \widetilde D_{
\delta }K{_{ab}}^{i}+ H^{ab}  D_{ \delta} \gamma_{ab}
 + H{^{abc}}_{i} \widetilde D_{\delta} (\widetilde
\nabla_{a}K{_{bc}}^{i}) ]d^{D} \xi,
\end{equation}
where
\begin{eqnarray}
\nonumber \\ H_{ab} \!\!\ & = & \!\! \frac{\partial L}{ \partial
\gamma_{ab}}, \nonumber \\
H{^{ab}}_{i} \!\!\ & = & \!\!\ \frac{\partial L}{\partial
K{_{ab}}^{i}} =H{^{ba}}_{i}, \nonumber \\
H{^{abc}}_{i} \!\!\! & = & \!\!\ \frac{ \partial L} { \partial
\widetilde \nabla _{a} K{_{bc}}^{i}} = H{^{acb}}_{i}.
\end{eqnarray}
On the other hand, using the deformation formalism introduced in
\cite{10} and writing the normal variation of $\gamma_{ab}$,
$K{_{ab}}^{i}$ and $\widetilde \nabla_{a} K{_{cb}}^{i}$ in a
curved background, we obtain
\begin{equation}
\widetilde D_{\delta} K_{ab}^{i} = - \widetilde \nabla_{a}
\widetilde \nabla_{b} \phi^{i} + K{_{ac}}^{i}K{^{c}}_{bj} \phi{j}
+ g(R(e_{a},n_{j})e_{b},n^{i}),
\end{equation}
\begin{eqnarray}
\nonumber \\ \widetilde D_{\delta} \widetilde
\nabla_{a}K{_{bc}}^{i} \!\!\! & = & \!\!\!  \widetilde
\nabla_{a}[-\widetilde \nabla_{b} \widetilde \nabla_{c} \phi^{i} +
K{_{db}}^{i}K{^{d}}_{cj} \phi^{j} +
g(R(e_{b},n_{j})e_{c},n^{i})\phi^{j} ]-[\widetilde
\nabla_{b}(K{_{a}}^{gj} \phi_{j}) + \widetilde \nabla_{a}
(K{_{b}}^{gj} \phi_{j}) \nonumber \\ \!\!\! & - & \!\!\!
\widetilde \nabla^{g} (K{_{ba}}^{j} \phi_{j})]K{_{gc}}^{i} +[-
\widetilde \nabla_{c}(K{_{a}}^{gj} \phi_{j}) - \widetilde
\nabla_{a}(K{_{c}}^{gj} \phi_{j} ) + \widetilde \nabla^{g}
(K{_{ca}}^{j} \phi_{j}) ] K{_{gb}}^{i} + [K{_{ad}}^{i} \widetilde
\nabla^{d} \phi^{j} \nonumber \\ \!\!\ & -  & \!\! K{_{ad}}^{j}
\widetilde \nabla^{d} \phi^{i} -
g(R(n_{k},e_{a})n^{j},n^{i})\phi^{k} ]K_{bcj},
\end{eqnarray}
with $g(R(e_{a},n_{j})e^{a},n^{i})= R_{\alpha \beta \mu \nu}
n{_{j}}^{\alpha} e{_{a}}^{\beta} e{^{a}}^{\mu} n{^{i}}^{\nu}$,
being $R_{\alpha \beta \mu \nu}$ the background Riemann tensor
\cite{3,10}.\\
Substituting  Eqs. (9), (10) and removing the scalar field
$\phi^{i}$ in (7) we obtain
\begin{eqnarray}
\nonumber \delta S \!\!\ & = & \!\!\ \int \sqrt{- \gamma}[K^{i}L -
2K^{abi} H_{ab} - \widetilde \nabla_{a} \widetilde \nabla _{b}
H{^{ab}}_{i} - K{_{ac}}^{i}K{_{b}}^{cj} H{^{ab}}_{j} +
g(R(e_{a},n^{i})e_{b},n^{j})) H{^{ab}}_{j} + \widetilde \nabla_{c}
\widetilde \nabla_{b} \widetilde \nabla _{a} H^{abci} \nonumber \\
\!\!\ & + & \!\!\ 2K_{{a}}^{gi} \widetilde \nabla_{b}
(H{^{abc}}_{j} K{_{gc}}_{j}) + 2K{_{b}}^{gi} \widetilde \nabla_{a}
(H{^{abc}}_{j} K{_{gc}}_{j} ) + K{_{ba}}^{i} \widetilde
\nabla^{g}(H{^{abc}}_{j}) K{_{gc}}^{j} + K_{a}{^{gi}} \widetilde
\nabla_{a} (H{^{abc}}_{j} K{_{gb}}^{j}) \nonumber \\
\!\!\ & - & \!\!\ K{_{ca}}^{i} \widetilde
\nabla^{g}(H^{abc}_{j}K{_{gb}}^{j}) - \widetilde \nabla
^{d}(H{^{abc}}_{j}K{_{ad}}^{j}K{_{bc}}^{i}) - \widetilde
\nabla^{d}(H^{abci} K{_{ad}}^{j}K_{bcj})-
g(R(n^{i},e_{a})n^{j},n^{l})H{^{abc}}_{l}K_{bcj} ] \phi_{i} \nonumber \\
\!\!\ & + & \!\!\ \int \sqrt{- \gamma} \widetilde \nabla_{a} [ L
\phi ^{a} - H{^{ab}}_{i} \widetilde \nabla_{b} \phi^{i} +
\widetilde \nabla_{b} H{^{ab}}_{i} \phi^{i} - H{^{abc}}_{i}
\widetilde \nabla _{b} \widetilde \nabla_{c} \phi^{i} +
H{^{abc}}_{i}g(R(e_{b},n_{j})e_{c},n^{i}) \phi^{j} + H{^{abc}}_{i}
K{_{db}}^{i} K{^{d}}_{cj} \phi^{j}  \nonumber
\\  \!\!\ & + & \!\!\  \widetilde \nabla _{b}
H{^{abc}}_{i} \widetilde \nabla_{c} \phi^{i} -\widetilde
\nabla_{b} \widetilde \nabla_{c} H{^{cba}}_{i} \phi^{i} -
H{^{bac}}_{i} K{_{gc}}^{i} H{_{b}}^{gj} \phi_{j} - 2 H{^{abc}}_{i}
K{_{gc}}^{i} K{_{b}}^{gj} \phi_{j} - H{^{cba}}_{i} K{_{gb}}^{i}
K{_{c}}^{gj} \phi_{j}  \nonumber \\
\!\!\ & + & \!\!\  H{^{gbc}}_{i} K{_{c}}^{ai} K{_{bg}}^{j}
\phi_{j} + H{^{gbc}}_{i} K{_{b}}^{ai} K{_{cg}}^{j} \phi_{j} +
H{^{gbc}}_{i} K{_{bcj}} K{_{g}}^{ai} \phi^{j} - H{^{dbc}}_{i}
K{_{d}}^{aj} K_{bcj} \phi^{i} ] d^{D} \xi,
\end{eqnarray}
from last equation we can identify the equations of motion given
by
\begin{eqnarray}
\nonumber  \!\!\ & K^{i} L  & \!\! - 2K{^{ab}}^{i} H_{ab} -
\widetilde \nabla_{a} \widetilde \nabla _{b} H{^{ab}}^{i} -
K{_{ac}}^{i}K{_{b}}^{cj} H{^{ab}}_{j} +
g(R(e_{a},n^{i})e_{b},n^{j})) H{^{ab}}_{j} + \widetilde \nabla_{c}
\widetilde \nabla_{b} \widetilde \nabla _{a} H^{abci} \nonumber \\
\!\!\ & + & \!\!\ 2K{_{a}}^{gi} \widetilde \nabla_{b}
(H{^{abc}}_{j} K{_{gc}}_{j}) + 2K{_{b}}^{gi} \widetilde \nabla_{a}
(H{^{abc}}_{j} K{_{gc}}_{j} ) + K{_{ba}}^{i} \widetilde
\nabla^{g}(H{^{abc}}_{j}) K{_{gc}}^{j} + K_{a}{^{gi}} \widetilde
\nabla_{a} (H{^{abc}}_{j} K{_{gb}}^{j}) \nonumber \\
\!\!\ & - & \!\!\ K{_{ca}}^{i} \widetilde
\nabla^{g}(H^{abc}_{j}K{_{gb}}^{j}) - \widetilde \nabla
^{d}(H{^{abc}}_{j}K{_{ad}}^{j}K{_{bc}}^{i}) - \widetilde
\nabla^{d}(H^{abci} K{_{ad}}^{j}K_{bcj}) \nonumber \\
\!\!\ & - & \!\!\ g(R(n^{i},e_{a})n^{j},n^{l})H{^{abc}}_{l}K_{bcj}
=0,
\end{eqnarray}
and we identify from the pure divergence term in (11)
\begin{eqnarray}
\nonumber \Psi^{a} \!\!\ & = & \!\!\  \sqrt{- \gamma} [ L \phi
^{a} - H{^{ab}}_{i} \widetilde \nabla_{b} \phi^{i} + \widetilde
\nabla_{b} H{^{ab}}_{i} \phi^{i} - H{^{abc}}_{i} \widetilde \nabla
_{b} \widetilde \nabla_{c} \phi^{i} +
H{^{abc}}_{i}g(R(e_{b},n_{j})e_{c},n^{i}) \phi^{j}  \nonumber
\\  \!\!\ & + & \!\!\  \widetilde \nabla _{b}
H{^{abc}}_{i} \widetilde \nabla_{c} \phi^{i} -\widetilde
\nabla_{b} \widetilde \nabla_{c} H{^{cba}}_{i} \phi^{i} -
H{^{bac}}_{i} K{_{gc}}^{i} K{_{b}}^{gj} \phi_{j} -  H{^{abc}}_{i}
K{_{gc}}^{i} K{_{b}}^{gj} \phi_{j} - H{^{cba}}_{i} K{_{gb}}^{i}
K{_{c}}^{gj} \phi_{j}  \nonumber \\
\!\!\ & + & \!\!\  2H{^{gbc}}_{i} K{_{c}}^{ai} K{_{bg}}^{j}
\phi_{j} + H{^{gbc}}_{i} K{_{bcj}} K{_{g}}^{ai} \phi^{j} -
H{^{dbc}}_{i} K{_{d}}^{aj} K_{bcj} \phi^{i} ],
\end{eqnarray}
as a symplectic potential for the theory described for a
Lagrangian given in equation (2), that is ignored in the
literature since that it does not contribute locally to the
dynamics, but generates our geometrical structure on the phase
space. Note that there exists a term involving explicitly the background curvature in Eq.(13).\\
Now, in the next lines we will take particular cases of the
Lagrangian given in equation (2) using the previous results  we
will confirm the results given in \cite{3}; for this, we
take as first example the [DNG] p-branes action. \\
As we know the [DNG] p-branes action is proportional to the area
of the spacetime trajectory created by the brane, thus, if we take
to $L=- \mu $, where $\mu$ is a constant characterizing the brane
tension we have the well known action for [DNG] p-branes
\begin{equation}
S= -\mu \int \sqrt{- \gamma} d^{D} \xi,
\end{equation}
in this manner, utilizing the Eqs. (8) we easily obtain
\begin{eqnarray}
\nonumber \\ H_{ab} \!\!\ & = & \!\! 0, \nonumber \\
H{^{ab}}_{i} \!\!\ & = & \!\!\ 0, \nonumber \\
H{^{abc}}_{i} \!\!\! & = & \!\!\ 0,
\end{eqnarray}
substituting the last result in the equation (12) we obtain
\begin{equation}
K^{i}=0,
\end{equation}
that corresponds to the equations of motion for [DNG] p-branes
describing extremal surfaces \cite{2,3,10}. \\
On the other hand, if we consider the Eq. (15) in (13) we find
\begin{equation}
\Psi^{a}= -\mu \sqrt{- \gamma} \phi^{a},
\end{equation}
that corresponds to the symplectic potential for [DNG] p-branes.
Thus, if we take the variation of $\Psi^{a}$ (the exterior
derivative on the phase space) given in (17) we will generate a
geometrical structure on the phase space, to more details see
\cite{3}.\\
As second example we will consider a Lagrangian that is quadratic
in the extrinsic curvature, because of  in many cases it was seen
that [DNG] action is inadequate and there are missing corrective
quadratic terms in the extrinsic curvature. For example, in the
eighties Polyakov proposed a modification to the [DNG] action by
adding a rigidity term constructed with the extrinsic curvature of
the worldsheet generated by a string, and  to include quadratic
terms in the extrinsic curvature to the [DNG] action is absolutely
necessary, because of its influence in the infrared region
determines the phase structure of the string theory, in this
manner, we can compute the critical behavior of random surfaces an
their geometrical and physical characteristics \cite{6}. In the
treatment of topological defects \cite{11}, curvature terms are
induced by considering an expansion in the thickness of the
defect. Bosseau and Letelier have studied cosmic strings with
arbitrary curvature corrections, finding for example, that the
curvature correction may change the relation between the string
energy density and the tension \cite{12}. Furthermore, such models
have been used to describe mechanical properties of lipid
membranes \cite{13}. Because of the above considerations, we will
take a Lagrangian quadratic in the extrinsic curvature given by
$L= \alpha K^{i}K_{i}$, here, $\alpha$ is a constant associated to
the brane tension \cite{3,10}. Thus, if we substitute it in Eq.
(8) we obtain
\begin{eqnarray}
\nonumber \\ H_{ab} \!\!\ & = & \!\! 2 \alpha K^{i} K_{abi}, \nonumber \\
H{^{ab}}_{i} \!\!\ & = & \!\!\  2 \alpha \gamma^{ab} K_{i}, \nonumber \\
H{^{abc}}_{i} \!\!\! & = & \!\!\ 0.
\end{eqnarray}
In this manner, in virtue to last equation the equation (8) takes the form
\begin{equation}
\widetilde \triangle K^{i}+
\left(-g(R(e_{a},n^{j})e^{a},n^{i})+(\gamma ^{ac}\gamma
^{bd}-\frac{1}{2}\gamma ^{ab}\gamma ^{cd}
)K{_{ab}}^{j}K{_{cd}}^{i}\right)K_{j}=0,
\end{equation}
that corresponds to the dynamics for the theory under study \cite{3,10}.\\
In the same form, if we substitute the Eq. (18) into (13) we
obtain
\begin{equation}
\Psi^{a}=2 \alpha \sqrt {-\gamma} \left[\frac{1}{2}K^{j}K_{j} \phi^{a} +
\phi_{i} \widetilde \nabla^{a}K^{i}- K_{i}\widetilde
\nabla^{a}\phi^{i}\right],
\end{equation}
that corresponds to the integral kernel of a covariant and invariant of gauge symplectic structure defined on
the covariant phase space \cite{3}. \\
To finish this section it is important to mention that in the same
form, using the previous results we can obtain the results
presented in \cite{4}; in this case, we analyze what happens when
we add the [GB] topological term to the [DNG] action in string
theory, and we found for example, that in the dynamics of
deformations exist a contribution non trivial  because of the [GB]
topological term, therefore, we found a contribution that does not
vanish in the symplectic structure constructed on the covariant
phase space for the [DNG-GB] system in string theory. These
important results allowed us find using a weakly covariant
formalism \cite{10}, the canonical variables for [DNG] p-branes
and [GB] term in string theory \cite{7}, however, as we already
commented we found some problems to consider the [DNG-GB] complete
system. In this manner, in the next section we will use a strongly
covariant formalism introduced in \cite{8} and the results
presented in \cite{9} for this problem can be clarified, in other
words,  we will find the canonical variables for [DNG-GB] system
in string theory, which is completely unknown in the literature.
\newline
\newline
\noindent \textbf{III. The canonical variables for DNG system  }\\[1ex]
As we commented, in \cite{7} we found the canonical variables for
[DNG] p-branes (that contain the particular case of strings
theory) using a weakly covariant formalism \cite{10},  with
spacetime indices, and the canonical variables for [GB]
topological term with worldsheet indices, in this manner, if we
consider the [DNG-GB] system we need rewrite the canonical
variables of [GB] term with background  spacetime indices and to
consider a canonical transformation that leaves the symplectic
structure in the Darboux form  with some new variables, $P$ and
$Q$, say. However, this problem can be clarified using a strongly
covariant formalism introduced
in \cite{8} as we will see in the next lines.\\
Using a strongly covariant formalism introduced by Carter
\cite{8}, it is found that the symplectic structure for [DNG]
branes in a curved background is given by \cite{9}
\begin{equation}
\omega= \sigma_{0}\int_{\Sigma} \delta(- \sqrt{-\gamma}
\eta{^{\mu}}_{\alpha}\xi ^{\alpha})d
\bar{\Sigma}_{\mu}=\int_{\Sigma} \sqrt{-\gamma} \widetilde J^{\mu}
d \bar{\Sigma}_{\mu},
\end{equation}
where $\sigma_{0}$ is a fixed parameter, $\eta{^{\mu}}_{\alpha}$
is the (first) fundamental tensor,  $\sqrt{-\gamma} \widetilde
J^{\mu} = \delta(- \sigma_{0}\sqrt{-\gamma}
\eta{^{\mu}}_{\alpha}\xi ^{\alpha})$, $\Sigma$ being a (spacelike)
Cauchy surface for the configuration of the brane while  $d
\bar{\Sigma}_{\mu}$ is the surface measure element of $\Sigma$,
and is normal to $\Sigma$ and tangent to the world-surface Here
$\delta$ is identified as exterior derivative on the covariant
phase space. The symplectic structure given in (21) is a exact
differential form, since it comes from the exterior derivative of
a one form and in particular is an identically closed two-form on
the phase space. The closeness is equivalent to the Jacoby
identity that Poisson brakets satisfy, in a usual Hamiltonian
scheme, and  the symplectic current is (world surface) covariantly
conserved ($\bar{\nabla}_{\mu} \widetilde J^{\mu}=0$), which
guarantees that $\omega$ is independent on the choice
of $\Sigma$ and, in particular, is Poncar\'e invariant.\\
We can rewrite the symplectic structure given in (25) for
identifying the canonical variables for [DNG] branes in the next
form
\begin{equation}
\omega=\int_{\Sigma} \delta X^{\alpha} \delta \hat{p_{\alpha}} d \Sigma,
\end{equation}
where $\hat{p_{\alpha}}=\sqrt{- \gamma} p_{\alpha}$, and
$p_{\alpha}= \sigma_{0} \tau_{\alpha}$, being $\tau_{\alpha}$ a
unit timelike vector field. In this manner, Eq. (32) allows us to
identify to $X^{\mu}$ and $\hat{p_{\alpha}}$ as the canonical
conjugate variables in this covariant description of the phase
space for [DNG] branes in a curved background (in \cite{7} we
identified the canonical variables for [DNG] p-branes in a weakly
covariant formalism). It is important to mention that the
symplectic structure given in equation (21) and the identification
of the canonical variables $X^{\mu}$ and $\hat{p_{\alpha}}$,
allows us to find for example, the covariant Poisson brackets, the
Poncar\'e charges and the closeness of the Poincar\'e algebra
\cite{7,9}.
\newline
\newline
\noindent \textbf{IV. The canonical variables for GB system in string theory }\\[1ex]
As we  know, the Einstein-Hilbert term is characterized for the
action
\begin{equation}
S= \sigma_{1} \int \sqrt{- \gamma} R d \bar{\Sigma},
\end{equation}
where $\sigma_{1}$ is a fixed parameter and $R$ is the scalar
curvature of the embedding \cite{5}. Using the deformations
formalism given in \cite{8} we can calculate the variation of $S$
obtaining
\begin{equation}
\delta S =2\sigma_{1}\int \sqrt{- \gamma}G^{\gamma \nu}K_{\gamma
\nu \mu}\xi^{\mu} d \bar{\Sigma} + \sigma_{1} \int \sqrt{-\gamma}
\bar{ \nabla}_{\mu}( -2 G{^{\mu}}_{\nu} \xi^{\nu}+\eta ^{\alpha
\beta} \delta \rho{_{\alpha}}{^{\mu}}_{\beta} -
\eta{^{\alpha}}_{\beta} \eta^{\mu \tau} \delta
\rho{_{\alpha}}{^{\beta}}_{\tau})d \bar{\Sigma},
\end{equation}
where $G^{\gamma \nu}$ is the internal adjusted Ricci tensor,
$K_{\gamma \nu \mu}$ is the second fundamental tensor and
$\rho{_{\alpha}}{^{\mu}}_{\beta}$ is  the frame gauge internal
rotation pseudo-tensor or internal connection \cite{8}. In
general, the adjusted Ricci tensor  does not vanish for a imbedded
$p$-surface, however, in string theory the Ricci tensor vanishes
identically. From last equation we can identify the equations of
motion for the brane theory given by
\begin{equation}
G^{\gamma \nu}K{_{\gamma \nu }}^{\mu} =0,
\end{equation}
and such as in the Sect. II,  the total divergence term of
equation (24) is identified as symplectic potential for the theory
under study, given by
\begin{equation}
\Psi^{\mu}= \sigma_{1} \sqrt{-\gamma} [-2 G{^{\mu}}_{\nu}
\xi^{\nu}+\eta ^{\alpha \beta} \delta
\rho{_{\alpha}}{^{\mu}}_{\beta} - \eta{^{\alpha}}_{\beta}
\eta^{\mu \tau} \delta \rho{_{\alpha}}{^{\beta}}_{\tau}].
\end{equation}
If we take the particular case of string theory in equation (25)
the adjusted Ricci tensor vanishes, in this manner, if we utilize
the standard canonical formalism to quantize this system, we would
not find apparently nothing interesting, however, as we can see in
\cite{4} using a weakly covariant formalism introduced in
\cite{10} we found that the [GB] term in string theory gives a
nontrivial contribution on the Witten covariant phase space
leading to a completely different quantum field theory. We can see
it if we takes the particular case of string theory in Eq. (26)
obtaining
\begin{equation}
\Psi^{\mu}= \sigma_{1} \sqrt{-\gamma} [\eta ^{\alpha \beta} \delta
\rho{_{\alpha}}{^{\mu}}_{\beta} - \eta{^{\alpha}}_{\beta}
\eta^{\mu \tau} \delta \rho{_{\alpha}}{^{\beta}}_{\tau}],
\end{equation}
in this manner, we can see that the terms of last equation do not
vanish. This result allows us to find
the canonical variables for [GB] strings.\\
In order to continue, we need rewrite the internal connection in
terms of the (co) vector $\rho_{\mu}$ defined as
\begin{equation}
\rho_{\lambda}= \rho{_{\lambda}}^{\mu}_{\nu} \varepsilon {^{\nu}}_{\mu}, \quad \rho{_{\lambda}}^{\mu}_{\nu}= \frac{1}{2}
\varepsilon{^{\mu}}_{\nu} \rho_{\lambda},
\end{equation}
where $\varepsilon {^{\mu \nu}} = 2 \iota_{0} ^{[\mu} \iota_{1}
^{\nu]}$, being $\iota_{0}^{\mu}$ a timelike unit vector, and
$\iota_{1}^{\mu}$ a spacelike one, which constitute an orthonormal
tangent (to the world sheet) frame \cite{8}. Thus, considering the
last equation, the symplectic potential given in the expression
(27) takes the form
\begin{equation}
\Psi^{\mu}= \sqrt{- \gamma} \varepsilon {^{\mu \nu}} \delta \rho_{\nu},
\end{equation}
where we have used the frame gauge property of $\rho{_{\lambda}}^{\mu}_{\nu}$ and consequently of $\rho_{\lambda}$ \cite{5}.\\
In this manner, we can define a covariant and gauge invariant
symplectic structure for [GB] strings as
\begin{equation}
\omega'= \int_{\sigma} \delta (\sigma_{1}\sqrt{- \gamma}  \varepsilon {^{\mu \nu}} \delta \rho_{\nu})d \bar{\Sigma}_{\mu},
\end{equation}
therefore, from last equation we can identify as well as for [DNG] system the canonical variables for [GB] strings, this is
\begin{equation}
p_{\nu} = \sigma_{1}\sqrt{-\gamma} \varepsilon {^{\mu }}_{\nu}
\tau_{\mu}, \quad q^{\nu}= \rho^{\nu}.
\end{equation}
In this manner, we can see that in this case the canonical
variables has spacetime indices contrary to \cite{7} that has
worldsheet indices. With these results we can treat the complete
[DNG-GB] system which  is the purpose of the next section and of
this paper.
\newline
\newline
\noindent \textbf{V. The canonical variables for DNG-GB system in string theory }\\[1ex]
In this section we will study the [DNG-GB] system in string
theory. For that, we begin with the action that describe the
system under study, this is
\begin{equation}
S=  - \sigma_{0} \int \sqrt{-\gamma} d \Sigma + \int \sigma_{1}
\sqrt{-\gamma} R d \Sigma,
\end{equation}
now, using the deformations formalism given in \cite{8}  we take
the variation of last equation and considering the particular case
of string theory, finding
\begin{equation}
\delta S= \sigma_{0} \int \sqrt{-\gamma} K_{\mu} \xi^{\mu} d
\Sigma + \int \bar \nabla_{\mu}[- \sigma_{0} \eta^{\mu}{_{\nu}}
\xi^{\nu} + \sigma_{1} \varepsilon^{\mu \nu} \delta \rho_{\nu}] d
\Sigma,
\end{equation}
where we can identify the equations of motion given by
\begin{equation}
 K ^{\mu}=0,
\end{equation}
that corresponds to the equations of motion for [DNG] strings. On
the other hand,  the total divergence term of Eq. (33) is
identified as symplectic potential for the theory under study
\begin{equation}
\Psi^{\mu}= \sqrt{- \gamma} [- \sigma_{0} \eta^{\mu}{_{\nu}}
\xi^{\nu} + \sigma_{1} \varepsilon^{\mu \nu} \delta \rho_{\nu}],
\end{equation}
Whit the previous results, form last equation we can  obtain the
covariant and gauge invariant  symplectic structure for [DNG-GB]
in string theory, this is
\begin{equation}
\omega= \int_{\Sigma} \delta \hat{P}_{\nu} \wedge \delta Q^{\nu}d
\bar{\Sigma},
\end{equation}
where
\begin{equation}
\hat{P_{\nu}}= \sqrt{- \gamma}p_{\nu}, \quad {\rm and} \quad
Q^{\nu}= -\frac{\sigma_{1}}{\sigma_{0}} \varepsilon^{\nu \alpha}
\rho_{\alpha} + X^{\nu},
\end{equation}
with $p_{\nu}= \sigma_{0} \tau_{\nu}$. Therefore, we can identify
to $\hat{P}_{\nu}$ and $Q^{\nu}$ as canonical variables for
[DNG-GB] system in string theory which is completely unknown in
the literature. We can note that the contribution because of [GB]
term  on the canonical variable $Q^{\nu}$ (see the first term of
$Q^{\nu}$ in Eq.(37)) will be relevant when is calculated the
angular momentum of
the complete [DNG-GB] system, and will be important in the complete quantum  field theory.\\
It is important to mention that we have choose the canonical
momentum for [DNG] strings (see the equation (22)) and [DNG-GB]
strings (see equation (37)) in the same form, the reason is that
$p_{\nu}$ satisfies the mass shell ($p_{\nu }p^{\nu}= \sigma^{2}$)
and as we know of the literature the  mass shell is an important
condition to quantum level for [DNG] strings, because of the
Virasoro operators and the mass shell conditions  determinate the
masses of the physical states,  in this manner, we also hope that
such condition  will be important when we analyze the spectrum of
[DNG-GB] system in string theory, however this we discuss in
future works.\\
To finish this work, is important to see that  if we take
$\sigma_{1}=0$ in equation (37) we obtain the result given in Eq.
(22). However, we hope  the choice of the canonical variables made
in this paper as first quantization are the best election, because
of the canonical momentum for [DNG] and [DNG-GB] strings coincide,
thus, with the results of this paper and  the treatment that is
found in the literature to quantize [DNG] strings we have the
necessary elements to quantize [DNG-GB] strings.
\newline
\newline
\newline
\noindent \textbf{V. Conclusions and prospects}\\[1ex]
\newline
As we can see, using the deformations formalism introduced by
Carter and a covariant and gauge invariant symplectic structure,
we could find the canonical variables for [DNG-GB] system in
string theory which is absent in the literature. Whit this results
we have the necessary  elements for study the quantization aspects
of [DNG-GB] strings, since for this purpose we need the results of
this paper and the  solutions to the equation of motion (34) that
are given in elementary books on string theory. In this manner, we
can observe the change on the resulting quantum field theory of
the topology of the world surface given by [GB] term, and thus,
find the contribution of such term on the results that we find in
the
literature for [DNG] strings; however, we will discuss this subject in future works. \\
In addition to this work, we know that the bosonic strings (which
is the case of this work) are not the general case to describe the
nature and it is necessary to add the supersymmetry, among other
things, in order to give a description of the fermionic matter, in
this manner, a interesting question may be the inclusion of
supersymmetry to the results of this paper to find the
quantization bases for [DNG-GB] in superstring theory an thus
giving
a complete description of the matter, however, we will discuss this  in future works. \\
\newline
\noindent \textbf{Acknowledgements}\\[1ex]
This work  was  supported by CONACyT under grant 44974-F .  The author wants to thank R. Capovilla for the support and the friendship that he has offered me. \\
\\[1em]

\end {document}